\documentclass[preprint2]{aastex}
\usepackage{graphicx}
\usepackage{amssymb}

\newcommand{\nav}{\ensuremath{\overline{n}}}

\begin{document}

\title{Giant Flare in SGR 1806-20\\
and Its Compton Reflection from the Moon}

\author{D.~D.~Frederiks\altaffilmark{1}, S.~V.~Golenetskii\altaffilmark{1},
V.~D.~Palshin\altaffilmark{1}, R.~L.~Aptekar\altaffilmark{1},
V.~N.~Ilyinskii\altaffilmark{1}, F.~P.~Oleinik\altaffilmark{1},
E.~P.~Mazets\altaffilmark{1}, and T.~L.~Cline\altaffilmark{2}}

\altaffiltext{1}{Ioffe Physico-Technical Institute, St.Petersburg,
194021, Russia; aptekar@mail.ioffe.ru}

\altaffiltext{2}{Goddard Space Flight Center, NASA, Greenbelt, MD
20771, USA}

\begin{abstract}
We analyze the data obtained when the Konus-Wind gamma-ray
spectrometer detected a giant flare in SGR~1806-20 on December 27,
2004. The flare is similar in appearance to the two known flares in
SGR~0526-66 and SGR~1900+14 while exceeding them significantly in
intensity. The enormous X-ray and gamma-ray flux in the narrow
initial pulse of the flare leads to almost instantaneous deep
saturation of the gamma-ray detectors, ruling out the possibility of
directly measuring the intensity, time profile, and energy spectrum
of the initial pulse. In this situation, the detection of an
attenuated signal of Compton back-scattering of the initial pulse
emission by the Moon with the Helicon gamma-ray spectrometer onboard
the Coronas-F satellite was an extremely favorable circumstance.
Analysis of this signal has yielded the most reliable temporal,
energy, and spectral characteristics of the pulse. The temporal and
spectral characteristics of the pulsating flare tail have been
determined from Konus-Wind data. Its soft spectra have been found to
contain also a hard power-law component extending to 10~MeV. A weak
afterglow of SGR 1806-20 decaying over several hours is traceable up
to 1~MeV. We also consider the overall picture of activity of
SGR~1806-20 in the emission of recurrent bursts before and after the
giant flare.
\end{abstract}

\keywords{neutron stars, flares, gamma rays, Compton scattering}

\section*{INTRODUCTION}
The first two soft gamma repeaters, SGR~0526-66
\citep{Mazets79a,Golenetskii84} and SGR~1900+14 \citep{Mazets79b},
were discovered and localized in March 1979. The third SGR~1806-20
was discovered in 1983 \citep{Atteia87,Laros87}. And only in 1998
was the fourth SGR~1627-41 discovered \citep{Woods99}. The situation
with the possible fifth SGR 1801-23 \citep{Cline00} arouses
scepticism, since only two soft bursts separated by an interval of
several hours have been detected from this source.

The emission of recurrent bursts by the gamma repeaters is highly
nonuniform in time. The gamma repeaters are predominantly in
quiescence. This phase can last for years, being interrupted by
reactivation periods that can be very intense.

The temporal and spectral characteristics for all of the above gamma
repeaters that have been observed over several years in the
Konus-Wind experiment are summarized in a unified catalog of SGR
activity \citep{Aptekar01}.

Giant flares, very rare events
comparable in peak emission power in the source ($\sim
10^{45}-10^{47}$~erg~s$^{-1}$) to the luminosity of quasars, are the
second, incomparably more impressive type of SGR activity.

The giant flare of March 5, 1979, had remained a unique event for
more than 19 years. On August 27, 1998, a giant flare came from
SGR~1900+14. All the main features of the flare in SGR~0526-66
manifested themselves in this flare: a narrow, very intense initial
emission peak with a hard energy spectrum accompanied by a
relatively weaker, spectrally soft tail that decayed for several
minutes while pulsating \citep{Mazets99a,Hurley99,Feroci99}. The
third similar, but even more intense flare that came from
SGR~1806-20 on December 27, 2004, was observed on many spacecraft
equipped with X-ray and gamma-ray detectors: INTEGRAL, Mars Odyssey,
Wind, Swift, RXTE, RHESSI, and others
\citep{Borkowski04,Hurley04,Mazets04,Palmer05,Hurley05,Smith05,Woods05}.

The enormous intensity of the initial pulse of the flare led to
detector overload and saturation. As a result, the pulse time
profile, spectrum, and intensity could not be measured reliably.
These characteristics have been estimated more reliably by analyzing
information from the small charged-particle detectors designed to
study low-energy plasma and mounted on the Geotail
\citep{Terasawa05}, RHESSI, and Wind \citep{Hurley05} spacecraft.

In this paper, we consider in detail the results obtained in our
Konus-Wind and Helicon-Coronas-F observations of the flare.

It should be noted that a peculiar giant flare was also observed in
SGR~1627-41 on June 18, 1998. It differed in characteristics from
the other three flares. The flare was a short single pulse with a
hard, rapidly evolving spectrum, but it had no pulsating tail and
was an order of magnitude less intense \citep{Mazets99b}.

A joint analysis of the four gamma repeaters allows us to point out
some features of their behavior on which we will dwell in the
subsequent discussion.

Thompson and Duncan (1995, 1996) hypothesized that the soft gamma
repeaters (SGRs) are young neutron stars with superstrong ($\sim
10^{15}$~G) magnetic fields that rapidly spin down due to the losses
through magnetodipole radiation.

\section*{INSTRUMENTATION}
The data considered here were obtained with the Konus gamma-ray
spectrometer onboard the NASA Wind spacecraft and with the Helicon
gamma-ray spectrometer onboard the Russian near-Earth Coronas-F
spacecraft.

\subsection*{Konus-Wind}
The Konus-Wind gamma-ray spectrometer was described in detail by
\citet{Aptekar95}.

For the convenience of the reader, let us briefly consider the main
design features of the spectrometer. Two spectrometric gamma-ray
detectors, S1 and S2, are placed on the spacecraft stabilized by
rotation around the axis perpendicular to the plane of the ecliptic.
Their axes are directed toward the south and north poles of the
ecliptic, respectively, which ensures an all-sky survey. The two
detectors operate independently of each other in two modes:
background and burst ones. In the background mode, each detector
measures the count rate with a resolution of 2.944~s in three energy
windows (G1, G2, and G3), with the results being directly written to
onboard memory. A trigger signal is generated when a burst is
detected by a particular detector at time T$_0$. This signal
triggers the burst mode in this detector. The burst time history is
recorded in the G1, G2, and G3 windows in the time interval from
T$_0$-0.512~s to T$_0$+229.632~s with a time resolution of 2, 16,
64, and 256~ms, which changes stepwise during the recording, and 64
multichannel spectra are measured in the energy range from 10~keV to
10~MeV with an accumulation time adapted to the current burst
intensity. The changes in the time resolution as the time history is
recorded are related to Konus telemetry capacity restrictions. The
measurement results are written to the instrument's random access
memory. On completion of the burst mode, the information is slowly
rewritten to onboard memory, which takes 1--1.5~h. For the rewriting
period, the instrument does not operate in the background mode.
Clearly, in this scheme of operation, there is a risk of losing
important information about the burst. Therefore, the instrument has
two standby measuring systems. The first system records portions of
the time history of a burst with a rapidly changing intensity of
duration from 0.128 to 8.192~s with a resolution of 2~ms and saves
information in random access memory if the recorded fast count rate
variations are statistically significant. The second system
continues to transmit the count rate measurements in the G2 window
during the rewriting over the service telemetry channel with a
resolution of 3.680~s. Note that both systems were used to record
the giant flare.

\subsection*{Helicon-Coronas-F}
The Helicon gamma-ray spectrometer is one of the instruments onboard
the Coronas-F solar space observatory \citep{Oraevskii02} that was
placed in a near-Earth low-eccentricity polar orbit (orbital
inclination $82\fdg5$, distance from the Earth 500--550 km) in June
2001. The spacecraft is stabilized by rotation around the axis
directed toward the Sun with a 10$^\prime$ accuracy.

Helicon is designed to record solar flares, gamma-ray bursts, and
SGR activity. The simultaneous operation of Helicon-Coronas-F and
Konus-Wind makes it possible to perform a comparative analysis of
the information about the same event from two instruments separated
in space by a distance from several hundred thousand to $\sim$2
million kilometers. This increases the reliability of the
observations of spectral and temporal features of the recorded
emission and, in many cases, allows a fairly accurate triangulation
of the sources to be performed. The Helicon gamma-ray spectrometer
is similar to the Konus spectrometer in characteristics of its two
detectors and in data presentation structure. One of its detectors
is oriented toward the Sun and the other detector scans the
antisolar hemisphere. The burst modes of operation of the two
spectrometers are similar. The background mode of operation of
Helicon was slightly modified. Eight energy windows in the range
10--200~keV are used to continuously monitor the hard X-ray emission
from the Sun with a time resolution of 1~s. Additionally, the solar
detector is equipped with a multichannel analyzer that continuously
measures the energy spectra in the range 200~keV--10~MeV with an
accumulation time of 32~s. The data are written to onboard memory
without any interruptions in the measurements. Recall, however, that
the usable exposure time during Coronas-F observations is severely
limited by the Earth's screening and by the satellite's passage
through the radiation belts at high latitudes and the South-Atlantic
Anomaly.

\section*{OBSERVATIONS}
The giant flare in SGR~1806-20 was detected by Konus-Wind on
December 27, 2004, in the burst mode, which was triggered by the
arrival of a short soft burst that led the flare by 143~s. The
record of these events in general form is presented in Fig.~1. The
appearance of a ``precursor'' displaced the flare into the time
resolution region of 0.256~s and reduced the time of its observation
by half. The related losses in the volume and detail of information
were largely compensated for by the operation of the Konus standby
systems mentioned in the Konus-Wind section.

\subsection*{The Precursor}
The time history of this burst is shown in Fig.~2a. The burst
emission is soft and is observed only in the G1 and G2 windows.
Figure~2b presents an averaged photon spectrum of the whole burst.
When it is fitted by the OTTB distribution, $kT$ is
36.9$\pm$0.9~keV. In its characteristics, the precursor does not
differ radically from other recurrent bursts in SGR~1806-20 while
slightly standing out by its intensity.

\subsection*{The Initial Pulse of the Flare}
A portion of the flare time history containing the giant initial
pulse is presented in Fig.~3. It was recorded by the first standby
system of the instrument in the G2 and G3 windows with a time
resolution of 2 ms. The flare began at T-T$_0$=142.98~s with an
avalanche-like increase in intensity that was traceable until the
detector's sharp transition to a state of deep saturation at
T-T$_0$=143.12~s. The enormous intensity of the flare emission
continued to keep the detector in a state of complete saturation for
0.50~s and only at T-T$_0$=143.62~s did the final segment of the
initial pulse decay become observable. Recall that when the giant
flare was detected by Konus-Wind on August 27, 1998, in SGR~1900+14
and when the data obtained were analyzed, it was established that
the Konus detector was deeply saturated at an incident flux of
$F_{E} \gtrsim 2.4 \times 10^{-2}$ erg~cm$^{-2}$~s$^{-1}$
\citep{Mazets99a}. For this flare, the saturation lasted for
$\simeq$0.2~s. All of the main flare evolution phases, its rise,
passage through the maximum, and decay, occurred in the time
interval when the detector was saturated. Undoubtedly, the actual
fluxes from the flare at these phases exceed the above instrumental
value of $F_E$ by many factors; the longer the saturation stage, the
greater this excess. However, further direct or indirect information
is required to reliably estimate the flare intensity. By a stroke of
luck, such direct information was obtained for the flare in
SGR~1806-20 during Coronas-F observations.

\subsection*{Compton Scattering of the Initial Pulse of the Flare by
the Moon}
During the flare of December 27, 2004, Coronas-F was shadowed from
SGR~1806-20 by the Earth. At 21:30:29.303 UT, Helicon recorded a
weak short burst whose profile is shown in Fig.~4a in the trigger
mode with a high time resolution of 2.048~ms. The burst lasted
$\simeq$180~ms. In this time, two multichannel spectra were measured
with an accumulation time of 65.536~ms each. The combined spectrum
is shown in Fig.~4b. The shape of the spectrum is quite unusual for
gamma-ray bursts. It was immediately suggested that the Compton
reflection of the giant initial pulse of the flare from the Moon was
observed. This suggestion was completely confirmed by examining the
spatial configuration of the Earth, the Moon, and the spacecraft
relative to the direction toward SGR~1806-20. At the flare time
(Fig.~5), the Sun was at an angular distance of $5\fdg3$ from
SGR~1806-20 and the Moon was located near the apogee of its orbit.
The full moon came on December 26. The angle through which the
gamma-rays of the initial pulse must be scattered in order to fall
on Coronas-F is $159\fdg5$--$159\fdg9$. The calculated time delay
between the arrivals of the flare at Wind and the scattered signal
at Coronas-F closely agrees with the observed value of 7.69~s. Thus,
the Helicon data make it possible to reliably determine the spectrum
and intensity of the initial pulse of the flare from the measured
spectrum of the scattered emission and to reconstruct the time
history of the most intense part of the flare from the profile of
the reflected signal. For this purpose, we numerically simulated the
scattering of a plane gamma-ray flux in the Moon's spherical surface
layer by the Monte-Carlo method using the GEANT4 software package
developed at CERN \citep{Agostinelli03}. The elemental composition
of the lunar soil was taken to be the following: O -- 42\%, Si --
21\%, Fe -- 13\%, Ca -- 8\%, Al -- 7\%, Mg -- 6\%, and the remaining
elements -- 3\%.

The response matrix of the Moon, which describes the escape
probability of a photon with energy $E^\prime$ scattered through
angle $\theta = 159^\circ$ normalized to a unit solid angle, was
obtained over a wide range of incident photon energies $E$, from
20~keV to 12~MeV. Figure 6a gives an idea of the matrix structure.
The intensity of filling the ($E^\prime$, $E$) plane with the dark
color is proportional to the escape probability of a photon with
energy $E^\prime$ for an incident photon with energy $E$. Two
regions of maximum probability can be distinguished. The dark curve
\begin{equation}
E' = \frac{E} {\left[1 + E / mc^2 (1 - \cos \theta)\right]}
\end{equation}
corresponds to single scattering through angle $\theta = 159^\circ$
with a limiting value of $E^\prime$=264~keV at high $E$. The
vertical line $E^\prime$=511~keV at $E > 2mc^2$ corresponds to the
escape of an annihilation photon as a result of the interaction of
photon $E$ with the production of an electron-positron pair. The
diffuse field in the graph characterizes the role of multiple
scattering, whose probability is high for a thick target. Figures 6b
shows examples of the matrix sections by the $E$=const plane, which
in each case represent the corresponding scattered photon energy
distribution function $f(E^\prime, E)$.

Using the response matrix, we can calculate the Moon's reflectances
in direction $\theta$ per unit solid angle in photon number
$\epsilon_N$ and energy $\epsilon_E$ normalized, for example, to one
incident photon with energy $E$ as a function of $E$. If, for
clarity, we pass from the summation over the matrix elements to
integration, then the expressions for the reflectances will take the
following form: $\epsilon_N(E) = \int_{E'_{min}}^{E'_{max}}f(E',
E)dE'$, $\epsilon_E(E) = (1/E)\int_{E'_{min}}^{E'_{max}}f(E',
E)E'dE'$. The lower limit $E'_{min}$=20~keV is determined by the
instrumental threshold. The upper limit $E'_{max}$=10~MeV was chosen
in the energy range where no scattered emission is observed at the
sensitivity level of the instrumentation. These dependences are
shown in Fig.~7.

The Moon's response matrix was then folded with the response matrix
of the Helicon gamma-ray spectrometer. To take into account the
lunar disk size $\pi R_L^2$ and the attenuation of the scattered
signal in its way to Coronas-F $\propto D_C^2$, the response matrix
was renormalized to the solid angle $\Omega = \pi R_L^2 / D_C^2$ at
which the Moon is seen from the satellite. In this form, the
response matrix of the Moon-Helicon system transforms the
differential photon spectrum of the flare $I(E)=A \, f(E)$ to the
energy loss spectrum measured by the gamma-ray spectrometer,
$n_i(E')$.

It follows from Fig.~4a that the reflected signal is weak. It is
distinguished above the background in the energy range
$\simeq$25--350~keV in the time interval T-T$_0$ from $\simeq$-40 to
$\simeq$140~ms. To improve the count statistics in the multichannel
spectrum, we grouped the spectrometer's channels by two and
considered the sum of the two spectra accumulated in the time
interval after T$_0$ spanning 32 4.096-ms time channels in the
signal profile. Since the limitations on statistics do not allow the
course of the fast spectral evolution to be traced on a millisecond
time scale in full, we will proceed from the assumption that the
shape of the flare spectrum $f(E)$ in this time interval
$\simeq$131~ms is retained and only its intensity $A$ changes. For
the subsequent analysis, note that the time profile of the total
signal in two windows in these 32 channels after the background
subtraction contains $363 \pm 22$ counts. The mean number of counts
per channel is \nav=11.3 in 4~ms. To model the instrumental
spectrum, we used the XSPEC v.11.3.2p code, including the derived
response matrix. We tested four models of the flare spectral
intensity $I(E)= Af(E)$: (a) a power law with an exponential cutoff,
(b) Band's model, (c) a power law, and (d) blackbody radiation.

The first model yields good agreement between the model and measured
spectra: $I(E)=A \, E^{-\alpha} \exp(-E/E_0)$. The values of the
parameters that minimize $\chi^2 = 10.6/12$~dof are
$A=1.98_{-0.41}^{+0.47} \times
10^6$~photons~cm$^{-2}$~s$^{-1}$~keV$^{-1}$, $\alpha =
0.73^{+0.47}_{-0.64}$, $E_0 = 666_{-368}^{+1859}$~keV; the energy of
the peak  $\nu F_\nu$ of the spectrum is $E_p =
850^{+1259}_{-303}$~keV. The errors in the parameters correspond to
a confidence probability of 90\%.

Band's model proved to be of little use for testing. At the poor
statistics in the hard part of the instrumental spectrum, only a
rough limit $\beta < -1.6$ can be placed on the index $\beta$ for
the high-energy power-law tail of the model fit. In this case, the
parameters $\alpha$ and $E_0$ do not differ from the previous model,
within the error limits, giving $\chi^2 = 10.3/11$~dof.

Formally, the power-law fit ($\gamma=-1.4\pm0.1$, $\chi^2 =
18.4/13$~dof) can be accepted, but it systematically overestimates
the soft part of the spectrum.

The assumption about blackbody radiation yields the worst result,
$\chi^2 = 27.4/13$~dof, $kT$=116~keV, and may be rejected.

The model fits for cases (a) and (d) are shown in Fig. 8.

The corresponding segment of the time history of the giant flare
that was completely lost on Konus-Wind when the detector was
saturated can be reconstructed from the time profile of the
reflected signal.

As we have noted above, the average spectral intensity of the flare
determined when the spectrum of the reflected signal was fitted,
$I(E) = A f(E)$~photons~cm$^{-2}$~s$^{-1}$~keV$^{-1}$, corresponds
to the mean number of counts \nav=11.3 in any time channel of the
profile $\Delta t=4.096$~ms. The intensity
$I_i(E)=(n_i/\overline{n}) A \,
f(E)$~photons~cm$^{-2}$~s$^{-1}$~keV$^{-1}$, will then correspond to
the number of counts $n_i$ recorded in channel $\Delta t_i$ in this
segment of the profile. We can also easily calculate the integrated
photon flux in interval $\Delta t_i$, $N_i(>E_i) = \int_E I_i(E)
dE$, and the integrated energy flux $F_i(>E_i) = \int_E E I_i(E)
dE$; the integration range extends from the sensitivity threshold of
the instrumentation $E_1 \simeq 20$~keV to the energy $E_2 \simeq
5$~MeV above which the contribution to the integral is negligible.

When reconstructing the time history of the initial pulse,
particularly its leading edge, we must take into account another
significant factor. When the flare is scattered by a spherical Moon,
the profile of the reflected signal is smeared. The plane flux from
the flare $N(t)$ reaches different areas of the lunar surface at
different times. Accordingly, the scattered photons come to a remote
observer with different delays. For backscattering, the maximum
difference in the relative delay is $\Delta \tau_{max} = 2 R_L/c =
11.6$~ms, where $R_L$ is the lunar radius and $c$ is the speed of
light. The relationship between the smeared profile of the signal
reflected from the Moon, $N_{refl}(t)$, and the flux from the flare
incident on it, $N(t)$, can be derived from simple geometrical
considerations:
\begin{equation}
N_{refl}(t) = k \int_0^2 N(t-\tau)(1-\tau/2) d \tau
\end{equation}
Here, the time $t$ and the delay $\tau$ are measured in units of
$R_L/c$. The coefficient $k$, which takes into account the
attenuation during scattering, is unimportant for the problem under
consideration and can formally be set equal to one.

This integral equation at the poor $N_{refl}(t)$ statistics can
hardly be effectively used to correct the entire profile $N(t)$.
However, the equation is very useful for estimating the width of the
leading edge of the flare. Our calculations indicate that the actual
width of the leading edge of the flare in the reflected signal
increases by $\leq 2 R_L/c = 11.6$~ms. Figure 9 shows the segment of
the time history of the initial pulse reconstructed from Helicon
observations in the interval $\Delta$T=180~ms. The fluence in it is
$S=0.87_{-0.24}^{+0.50}$~erg~cm$^{-2}$ at a confidence level of
90\%. Outside this interval, the assumption that the average shape
of the spectrum $f(E)$ is retained is no longer applicable. On the
contrary, the G2/G1 hardness variation shown in Fig. 4 indicates
that the flare spectrum rapidly becomes increasingly soft. Note that
such a situation took place in the August 27, 1998 flare from
SGR~1900+14 (Mazets et al. 1999a). The fluence in the decaying tail
of the pulse does not exceed a few percent of the fluence in the
interval $\Delta$T=180~ms. Thus, the estimate of
$S=0.87_{-0.24}^{+0.50}$~erg~cm$^{-2}$ characterizes the fluence of
the entire pulse, within the error limits.

\section*{THE PULSATING TAIL OF THE FLARE}
The Konus-Wind record of the pulsating tail of the flare in three
energy windows (G1, G2, and G3) is shown in Fig. 10. Also shown is
the time variation of the hardness ratio G2/G1. The count rate in
the G3 window is low, but it exceeds the background level
statistically significantly. After T-T$_0$=230~s, the record of the
flare ended, giving way to data output to onboard memory. The count
rate in the G2 window recorded with a resolution of 3.68~s by the
second standby system of the instrument gives a general idea of the
duration and intensity of the flare tail. The accumulation time,
3.68~s, is close to half the pulsation period, 7.56~s, which leads
to the pattern of deep beats. Therefore, a record with a resolution
of 7.36~s is shown in Fig. 11.

The energy spectra in the flare tail in the segment of transition
from the main peak to the steady-state pattern of pulsations until
T-T$_0$=152~s was measured with an accumulation time of 0.256~s.
Subsequently, at the final stage of the trigger mode of
measurements, the accumulation time was 8.192~s.

The soft part of the photon spectra, just as the spectra of
recurrent bursts, is well fitted by a distribution close to OTTB
radiation. Figure 12 shows the time history of the initial part of
the pulsating tail in the G1 and G2 windows and gives the spectral
parameters $kT$ for 30 spectra measured with an accumulation time of
0.256~s. We see that $kT$ correlates with the radiation intensity.

The hard part of the pulsation spectra exhibits a significant
feature that was not noticed previously in two other flares: at
energy $\simeq$200~keV, the exponential cutoff with $kT \simeq 30$
transforms into a hard power-law tail that extends with an index
$\gamma \simeq -1.7$ to energy $\sim$10~MeV. The intensity of the
emission in the tail is low and can be detected with an acceptable
statistical accuracy by adding up several successive spectra. As an
example, Fig. 13 presents several such spectra. The time variation
of the energy flux in the power-law tail at energies 0.5--8~MeV is
shown in Fig. 14. Unfortunately, the data obtained do not allow the
question of whether the intensity of the power-law tail is modulated
with a period of 7.56~s to be clarified due to the long accumulation
time of the spectra.

Under the assumption that the shape of the average spectrum shown in
Fig. 13d does not change significantly throughout the decaying tail
of the flare (Fig. 10), the fluence in it is $S_{tail} \simeq 8
\times 10^{-3}$~erg~cm$^{-2}$.

\section*{THE SGR AFTERGLOW}
Yet another feature in the behavior of SGR~1806-20 was observed
after the flare. \citet{Mereghetti05} reported that the SPI-ACS
system onboard the INTEGRAL satellite observed the appearance and
decay a hard afterglow from SGR~1806-20 with a fluence comparable to
the fluence in the flare tail in the interval from $\sim$400 to
$\sim$4000~s after the onset of the giant flare.

The Konus-Wind measurements in the background mode were resumed
after the flare data were rewritten to memory 5075~s after T$_0$.
Nevertheless, we analyzed the background measurements a day before
and after the flare; the results are presented in Fig. 15.
Generally, the background variations in the hard G2 and G3 windows
are determined by the flux variations of cosmic ray, mostly of a
solar origin. The Z ($E>10$~MeV) window serves to monitor the
cosmic-ray flux. No data in the soft G1 window were used, since
unpredictable variations in the flux from X-ray sources manifest
themselves in it.

Note that SGR~1806-20 lies near the ecliptic and the two Konus-Wind
detectors, S1 and S2, are irradiated by the source identically. The
observed small differences in the count rates are related to the
actual boundaries of the energy windows.

We see from Fig. 15a that the background variation in the G2 and G3
windows of both detectors generally follows the background in the Z
window. However, immediately after the interruption of the
measurements to rewrite the data, excess emission without any
signatures in Z was observed in the G2 and G3 windows. This picture
is consistent with the assumption that the source of hard X-ray and
gamma-ray emission began to act during the interruption of the
measurements and was located near the ecliptic. We may conclude that
the decaying phase of the SGR afterglow detected on INTEGRAL is
observed. The data from the two detectors were added and the results
are presented in Fig. 15b. The afterglow is traceable until
T-T$_0\sim$12000~s. The ratio of the sums of the counts in the G3
and G2 windows is 0.385$\pm$0.055. For a power-law spectrum, this
hardness ratio corresponds to an index of $\sim -1.6$. The fluence
in the observed phase of the afterglow in the energy range
80--750~keV is estimated to be $\sim 2 \times
10^{-4}$~erg~cm$^{-2}$.

\section*{THE SGR ACTIVITY BEFORE AND AFTER THE FLARE}
The emission of recurrent bursts from SGR~1806-20, just as from
other gamma repeaters, is distributed very nonuniformly in time.
SGR~1806-20 was detected and localized in the period of high
activity in the early 1980s \citep{Atteia87,Laros87}. The source was
also highly active in 1996. From January to May 2004, Konus-Wind
observed a total of two bursts in the trigger mode. However, a new
period of high activity began in May and Konus and Helicon recorded
74 trigger bursts before the giant flare. After the flare until the
end of 2005, 22 events were observed. These data were plotted in the
$S$--$\Delta$T diagram (Fig. 16). The 2005 events are slightly less
intense (by a factor of 2--3). The precursor, a short burst appeared
142~s before the flare, is also presented in the diagram. It is
strongest in energetics, but it should hardly be considered as an
event that differs radically from the entire set. In our opinion
\citep{Golenetskii04}, the appearance of close groups or series of a
larger number of recurrent bursts filling a time interval of several
minutes is considerably stronger evidence for the preflare state of
the source. Three such series were observed in SGR~1806-20: on
October 5, 2004 (83 days before the flare), December 21, 2004 (6
days before the flare), and December 25, 2004 (2 days before the
flare). It is important to mention that a similar series of
recurrent bursts was observed on May 30, 1998, in SGR~1900+14 89
days before the giant flare of August 27, 1998 \citep{Aptekar01}.
Note also that the Sun was highly active in the summer of 1998. For
many hours and days, the background level attributable to intense
solar cosmic-ray fluxes often increased to a level that ruled out
the trigger recording of both single short bursts and, possibly, a
new series.

All four series are presented in Fig. 17 in the record with a time
resolution of 0.256~s. The total fluence of each series $S$ is given
in the caption to the figure.

\section*{DISCUSSION}
The time history of the activity of SGR~1806-20 in 2004 considered
here, whose culmination was the giant flare on December 27, 2004,
can be divided into a series of characteristic stages.

\begin{enumerate}
\renewcommand{\labelenumi}{(\arabic{enumi})}

\item The source's reactivation accompanied by an enhanced
emission rate of recurrent bursts with cases of their close grouping
into short series of several tens of events.

\item The emission of a giant initial pulse of the flare with a
hard, rapidly evolving spectrum.

\item The stage of transition from the initial pulse to the
pulsating tail of the flare.

\item The decaying soft pulsating tail of the flare.

\item The prolonged hard afterglow of the source.

\item The gradual decay of the emission of recurrent bursts among
which separate longer-duration events appear.

\end{enumerate}

Such a sequence of events, at minimum deviations from it, was mainly
also characteristic of the development of activity in SGR~1900+14 in
1998 with the giant flare on August 27, 1998 \citep{Mazets99a}. In
both cases, the observations were performed with the same Konus-Wind
instrument. However, late in August, the solar activity was high and
the solar cosmic-ray fluxes led to a manifold increase in the
radiation background level. The high background prevented the
detection of such weak effects as the remote afterglow of the source
or the hard powerlaw tails in the pulsation spectrum. For this
reason, the question of whether these exist for the flare in
SGR~1900+14 is still an open question.

The giant flare of March 5, 1979, in SGR~0526-66 and the subsequent
emission of recurrent bursts until the middle of 1983 were observed
on Venera space stations with the Konus instrumentation
\citep{Aptekar01}. This instrumentation was appreciably inferior in
sensitivity and, particularly, information content to the Konus-Wind
instrumentation. Nevertheless, it can be assumed with a fair degree
of confidence that SGR~0526-66 was in quiescence before the flare of
March 5, 1979, in half a year of observations. Note that this gamma
repeater, having emitted $\simeq$20 recurrent bursts, has remained
quiescent for more than 20 years.

Thus, we may conclude that, despite the close similarity between the
giant flares in the three SGRs, the individual differences in their
activity before and after the flare show up quite clearly. Moreover,
the universality of the picture of the giant flare itself in SGR is
violated by the very intense flare of June 18, 1998, in SGR~1627-41
\citep{Mazets99b}. The flare was several thousand times more intense
than the recurrent bursts in this gamma repeater. This flare lasted
$\simeq$0.5~s, but it was abruptly cut off, without leaving even the
slightest trace of the decaying tail.

\subsection*{The Picture of Pulsations}
It is of interest to compare the pulsations in the three flares.
Their profiles averaged over several periods to smooth out small
fluctuations are shown in Fig. 18. Each profile consists of several
(from two to four) overlapping peaks. In all three cases, the
maximum modulation depth is approximately the same, $\simeq 85\%$.

The flare tail is widely believed to be emitted by a cloud of dense
relativistic electron-positron plasma with an admixture of baryons
trapped and confined by the superstrong magnetic field of a neutron
star when the fireball with which the giant initial pulse of the
flare is associated is ejected from it. The ultimate configuration
of the trapped part of the fireball is formed during the transition
from a short initial pulse to a uniform repetitive picture of
pulsations. The rigid fixing of the magnetic trap's field lines in
the neutron star's highly conductive crust ensures the subsequent
stability of the formed cloud configuration.

In our opinion, the crucial role in forming the observed picture of
deep modulation of the slowly decaying flare tail emission belongs
to the shape of the angular beam and to its changing (due to the
rotation of the neutron star) orientation relative to the observer's
position on the celestial sphere. In the inertial equatorial
coordinate system associated with the neutron star, the observer's
position is specified by right ascension $\alpha_{obs}$ and
declination $\delta_{obs}$. For the emission to be recorded as a
separate peak, it must have a beam that crosses the observer's
celestial parallel, a small circle with $\delta = \delta_{obs}$,
and, hence, must pass through the observer's direction as the star
rotates. In this case, the observed shape of the peak will be
determined by the shape of the angular beam in its section by the
$\delta_{obs}$ parallel. The angular beam FWHM in this section
$\Delta \theta$ is related to the apparent FWHM of the peak in the
star's rotation phase $\Delta \phi$ by $\Delta \theta = 2 \pi \Delta
\phi \cos \delta_{obs}$. When considering the shape of the peak
recorded in different energy windows (see, e.g., Fig. 10), we can
establish that the beam width along the declination circle decreases
with increasing energy. The overall variation in hardness G2/G1 and
its peculiarities in the region of overlap between the neighboring
pulsation peaks can be attributed precisely to this circumstance.

However, when the structure of the pulsations with several peaks is
considered, the difficult question of the number of emission sources
on the neutron star, including the sources invisible to the
observer, arises.

The assumption about the distribution of trapped plasma in large
number of magnetic traps would be an undesirable complication of the
flare tail emission model. The situation becomes slightly less acute
if we assume that the angular beam is close to a fan pattern, having
the shape of a widely opened hollow cone. At each given time, such a
beam will be projected onto the celestial sphere in the form of a
long stripe elongated along a celestial circle of a large angular
radius inclined to the stellar equator. Such a stripe can cross the
observer's parallel $\delta_{obs}$ twice and two pulsations peaks
will be observed in one rotation of the neutron star, as in
SGR~0526-66. If no additional, more peculiar assumptions about the
complex spatial shape of the beam is invoked, then the presence of
two fan beams will be required to explain the pattern of pulsations
in SGR~1900+14 and SGR~1806-20. The reasoning behind the formation
mechanism of two fan beams given ad hoc for the flare of August 27,
1998, \citep{Thompson01} does not look quite convincing, but it
undoubtedly serves as the first evidence for the existence of yet
another problem requiring its solution. To explain the peculiarities
of the giant flare in SGR~1627-41, we can assume that, in this case,
the direction of the neutron star's rotation axis makes a small
angle with the observer's line of sight. The flare occurred in the
stellar hemisphere invisible to the observer. The magnetic trap with
trapped plasma remains invisible at all rotation phases of the star
and the possibility of detecting the flare tail is completely ruled
out.

\subsection*{The Energetics of SGR~1806-20}
The fluence of the initial pulse of the flare at energies $E>
16.5$~keV in the time interval after the flare onset $\Delta$T=0.6~s
is $S=0.87_{-0.24}^{+0.50}$~erg~cm$^{-2}$. The peak flux of the
flare in a time interval of 4~ms is
$F_{peak}=13.1_{-4.4}^{+8.0}$~erg~cm$^{-2}$~s$^{-1}$. The fluence in
the transition region and the pulsating tail of the flare before its
decay is $S_{tail} \simeq 8 \times 10^{-3}$~erg~cm$^{-2}$. The flux
at the main pulsation peak for the first rotations of the star is $F
\simeq 4.9 \times 10^{-5}$~erg~cm$^{-2}$~s$^{-1}$ (in 0.256~s). The
intensity of recurrent bursts varies over a wide range, from $\sim 2
\times 10^{-7}$ to $\sim 2 \times 10^{-4}$~erg~cm$^{-2}$ at a
duration of $\sim$0.1--20~s (Fig. 16).

The estimates of the distance $D$ to SGR~1806-20 are being discussed
(see, e.g., \citet{Cameron05,McClure05}). We take the widely used
value of $D$=15~kpc. We also assume that the emission of the initial
pulse is isotropic in a solid angle of $4 \pi$~sr. Our estimates of
the energy output and peak luminosity of the initial pulse, tail,
and recurrent bursts are summarized in the Table~1.

For the afterglow, we estimated the energy release only in its final
stage. According to \citet{Mereghetti05}, the total energy of the
afterglow is comparable to the energy emitted in the flare tail.

\acknowledgements{This work was supported by the Federal Space
Agency of Russia and the Russian Foundation for Basic Research
(project no. 06-02-16070).}


%

\vspace{1cm}
\begin{flushright}
\textit{Translated by V. Astakhov}
\end{flushright}

\onecolumn
\begin{deluxetable}{rcc}
\tablecaption{Energetics of SGR~1806-20}
\tablehead{\colhead{} & \colhead{$Q_{rad}$, erg} &
\colhead{$L_{max}$, erg s$^{-1}$}}
\startdata
Initial pulse & $\simeq 2.3 \times 10^{46}$ & $\simeq 3.5 \times 10^{47}$ \\
Pulsating tail & $\simeq 2.1 \times 10^{44}$ & $\simeq 1.3 \times 10^{42}$ \\
Recurrent bursts & $5\times 10^{39}$--$5 \times 10^{42}$ & $2\times 10^{41}$--$2 \times 10^{42}$ \\
\enddata
\end{deluxetable}

\clearpage
\begin{figure}
\centering
\includegraphics[width=\textwidth]{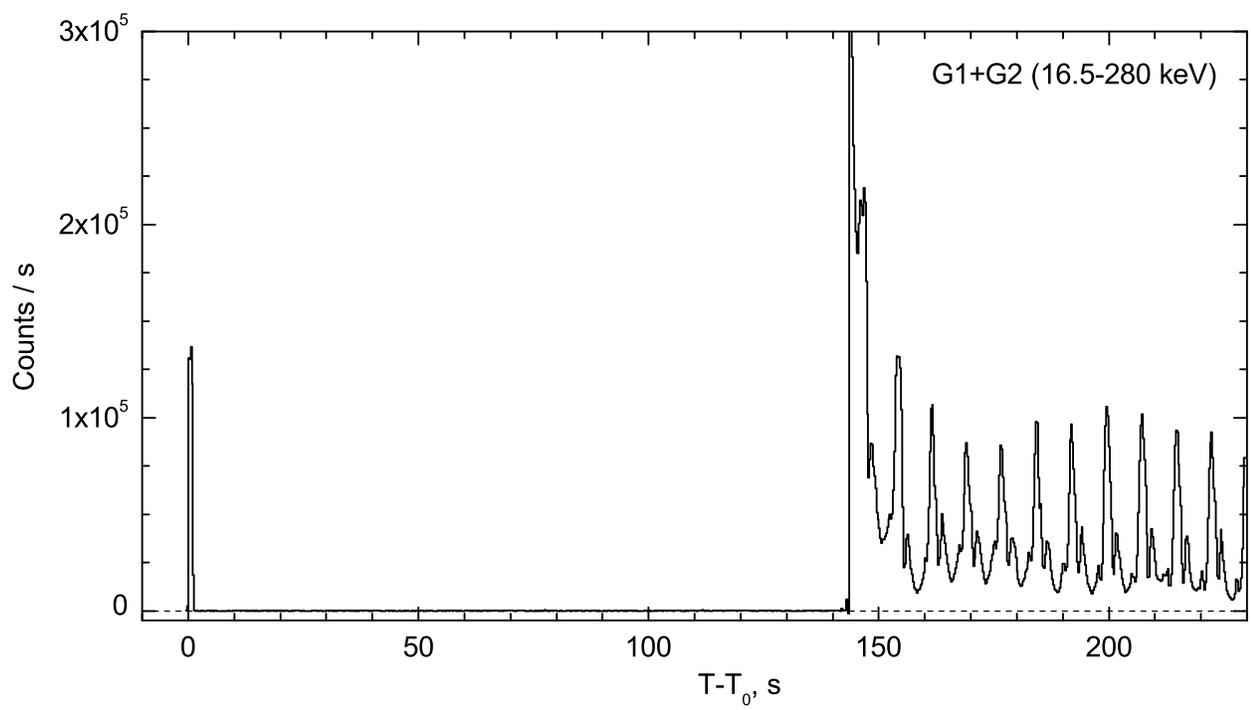}
\caption{Trigger record of the December 27, 2004 flare for the sum
of the G1 and G2 windows with a time resolution of 0.256~s.
\label{tKW_giant_and_precursor}}
\end{figure}
\clearpage
\begin{figure}
\centering
\includegraphics[width=\textwidth]{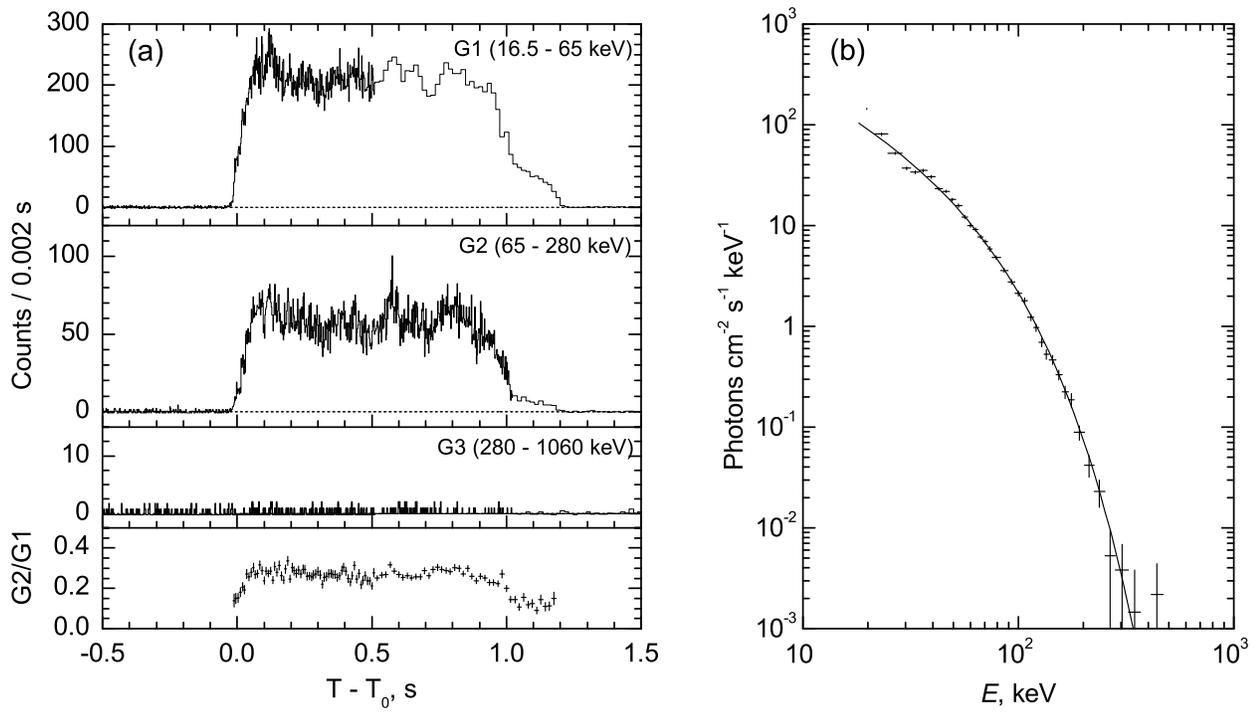}
\caption{Precursor of the giant flare. (a) The time history in three
energy windows recorded by the Konus-Wind detector at
T$_0$=21:27:59.447 UT (the background was subtracted). In the G2 and
G3 windows, the resolution is 2 and 16~ms before and after 1~s,
respectively; in the G1 window, the resolution is 2 and 16~ms before
and after 0.512~s, respectively. (b) The spectrum measured in
T-T$_0$=0--1.024~s. \label{tKW_giant_precursor}}
\end{figure}
\clearpage
\begin{figure}
\centering
\includegraphics[]{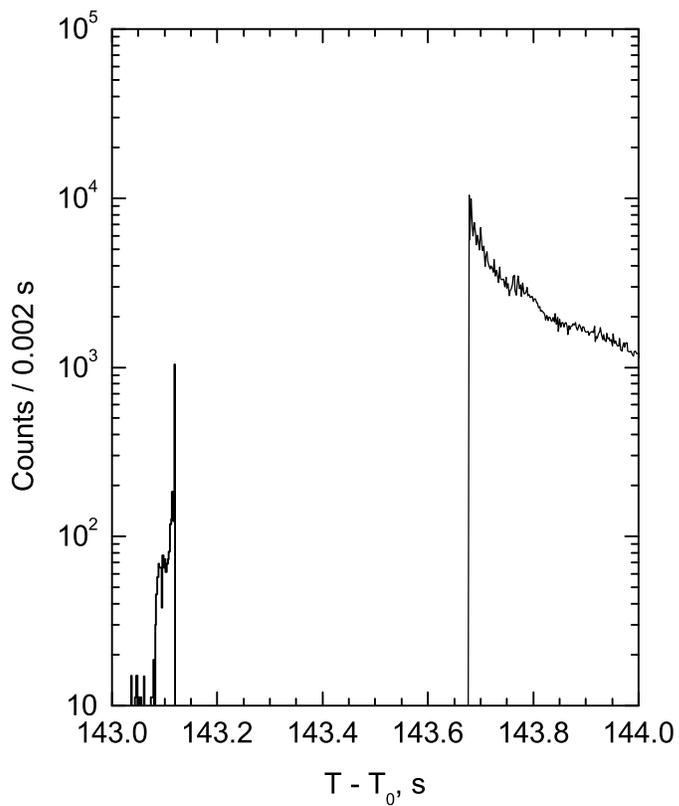}
\caption{Initial pulse of the giant flare (the sum of windows
G2+G3). The sharp boundaries of the state of complete saturation are
indicative of an enormous rate of change in emission intensity at
the leading edge and decay of the flare. \label{tKW_giant_initial}}
\end{figure}
\clearpage
\begin{figure}
\centering
\includegraphics[width=\textwidth]{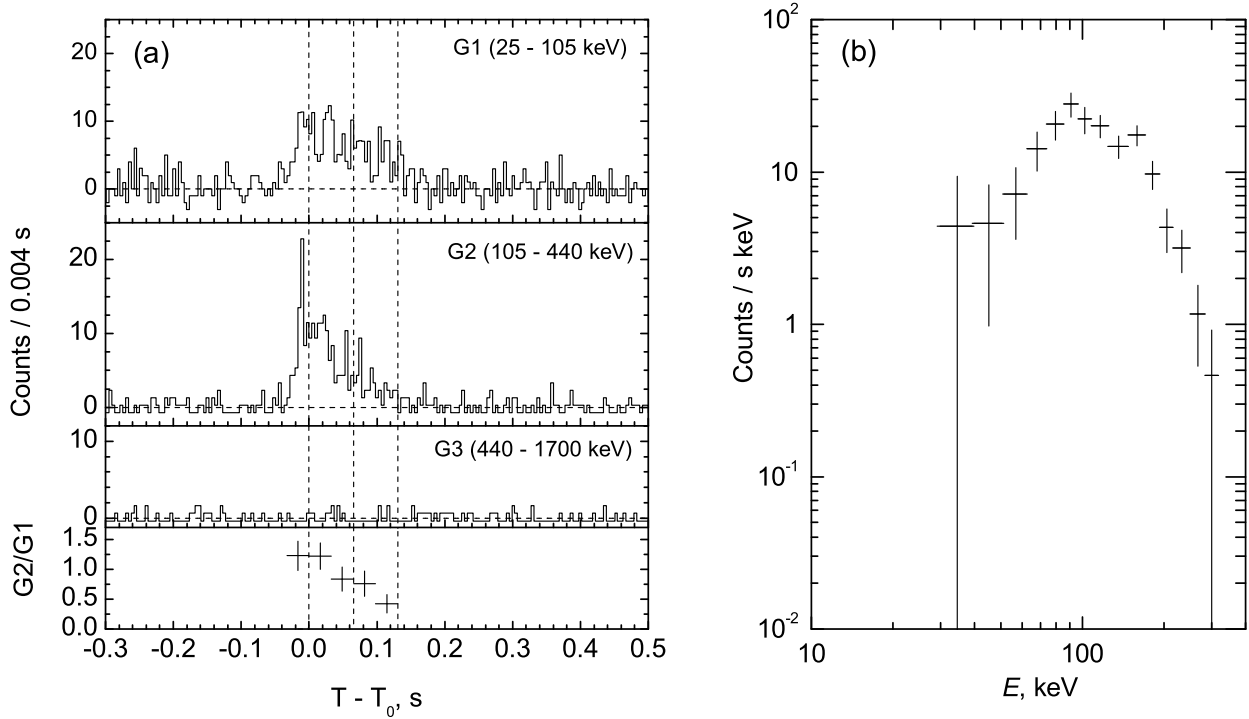}
\caption{(a) Helicon time history of the initial pulse of the giant
flare on December 27, 2004, reflected from the Moon.
T$_0$=21:30:29.303 UT, the time resolution is 4 ms, the background
was subtracted. The hardness (G2/G1) variation points to rapid
spectral evolution of the emission. The vertical lines mark two
measurement intervals of multichannel spectra. (b) The combined
spectrum for the two intervals. \label{tHC_giant_reflected}}
\end{figure}
\clearpage
\begin{figure}
\centering
\includegraphics[width=\textwidth]{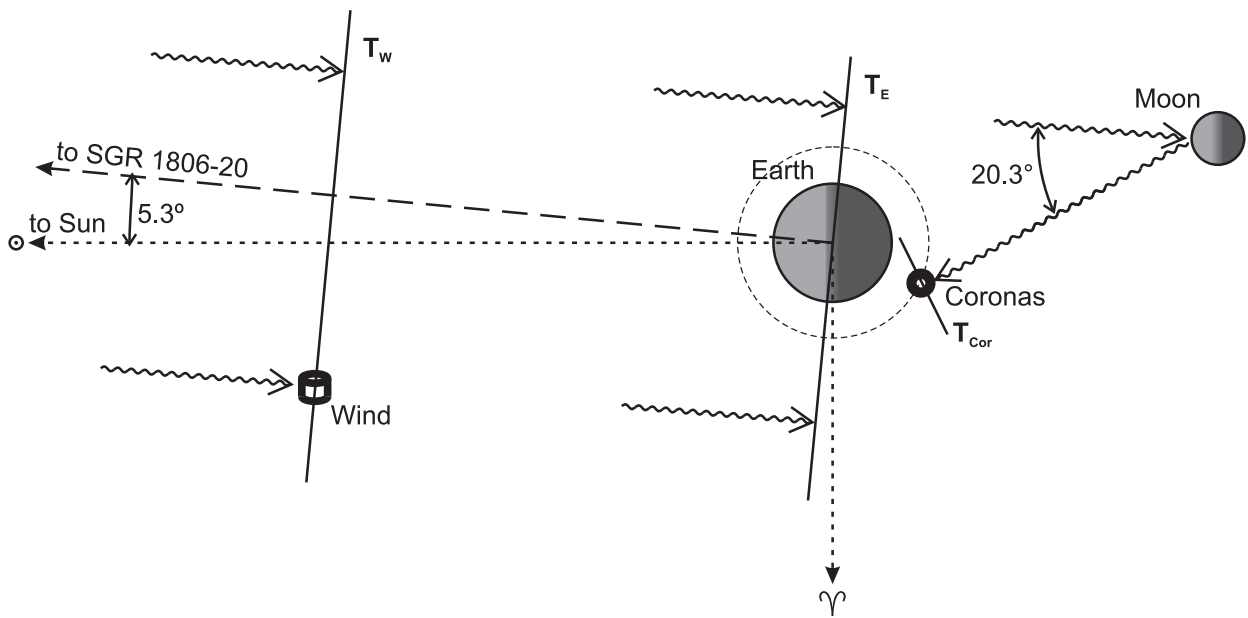}
\caption{Scheme illustrating the Konus-Wind and Helicon-Coronas-F
observations of the giant flare. The leading edge of the flare from
SGR~1806-20 arrives at Wind at time T$_W$, passes by the Earth at
T$_E$=T$_W$+5.086~s, reaches the Moon and is reflected from it, and,
finally, the reflected emission reaches the Helicon-Coronas-F
detector at T$_{Cor}$=T$_W$+7.69~s.  \label{Giant_scheme}}
\end{figure}
\clearpage
\begin{figure}
\centering
\includegraphics[width=\textwidth]{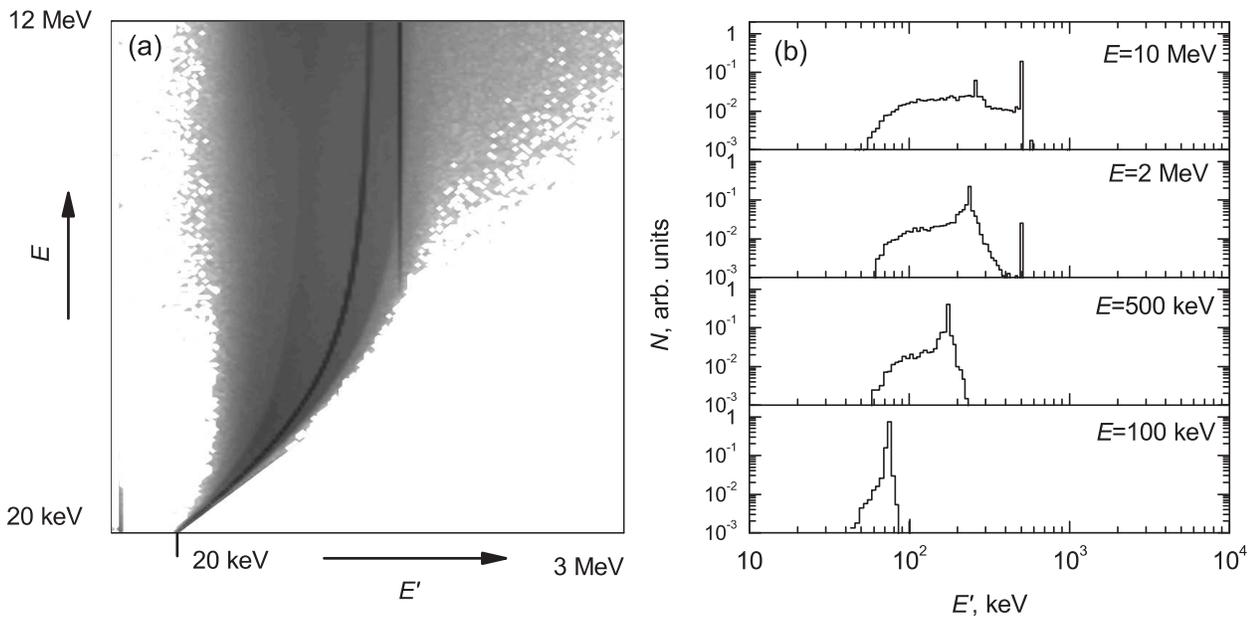}
\caption{Response matrix of the Moon calculated for an angle of
reflection of 159$^\circ$. (a) The quasi-three-dimensional
distribution of escaped photons for various energies $E$ of incident
$\gamma$-ray photons: the darker color corresponds to a higher
escape probability of a photon with given energy $E^\prime$. The two
dark lines are the curve corresponding to single Compton scattering
and the straight line characterizing the escape of annihilation
0.511-MeV photons due to the production of pairs by hard flare
photons. (b) The energy distribution of escaped photons for several
$E$. \label{MoonResponseMatrix}}
\end{figure}
\clearpage
\begin{figure}
\centering
\includegraphics[]{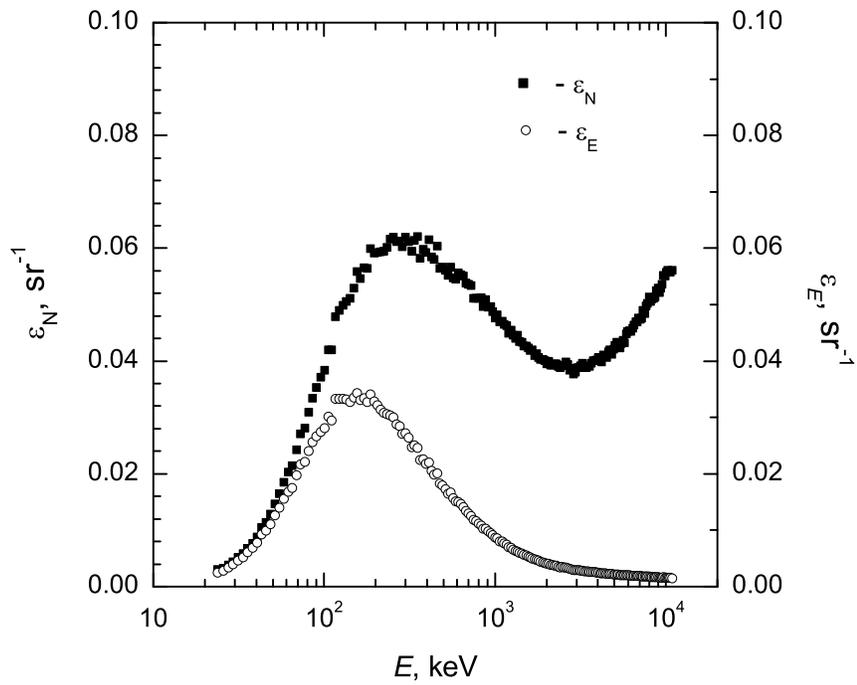}
\caption{Efficiency of the reflection of $\gamma$-ray emission by
the Moon in photon number $\epsilon_N$ and energy $\epsilon_E$:
respectively, the number and energy carried away by the photons
escaped at angle $\theta = 159^\circ$ per unit solid angle $\Omega$
as a function of $E$. \label{MoonEfficiency}}
\end{figure}
\clearpage
\begin{figure}
\centering
\includegraphics[width=\textwidth]{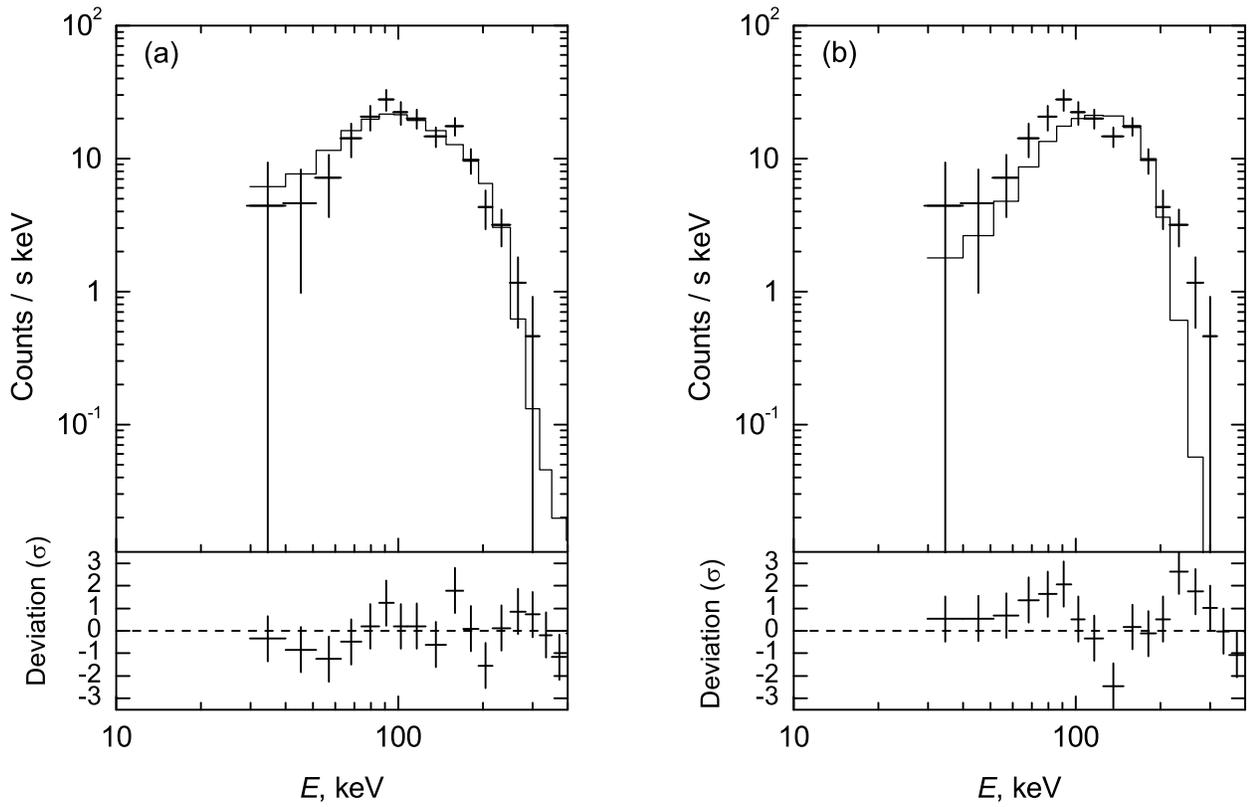}
\caption{Combined spectrum of the reflected initial pulse of the
giant flare recorded by the Helicon detector (accumulation time
0.131 s). The solid step lines indicate the various spectral model
fits. (a) A power law with an exponential cutoff,
$\chi^2$=10.6/12~dof; (b) a blackbody model,  $\chi^2$=27.4/13~dof.
\label{spHC_giant_reflected}}
\end{figure}
\clearpage
\begin{figure}
\centering
\includegraphics[]{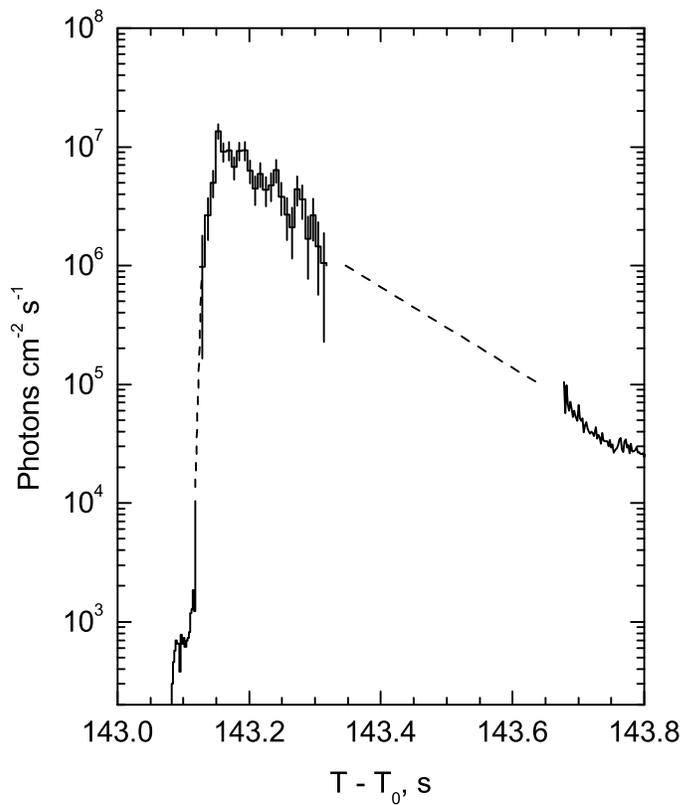}
\caption{Reconstructed time history of the initial pulse of the
giant flare. The upper part of the graph was reconstructed from
Helicon data (1$\sigma$ statistical errors are shown) and the lower
part was reconstructed from Konus-Wind data. The dashed lines denote
the intervals in which the emission is intense enough to saturate
the Konus-Wind detector, but is not intense enough for the reflected
signal to be recorded by Helicon. \label{tGiant_reconstructed}}
\end{figure}
\clearpage
\begin{figure}
\centering
\includegraphics[width=\textwidth]{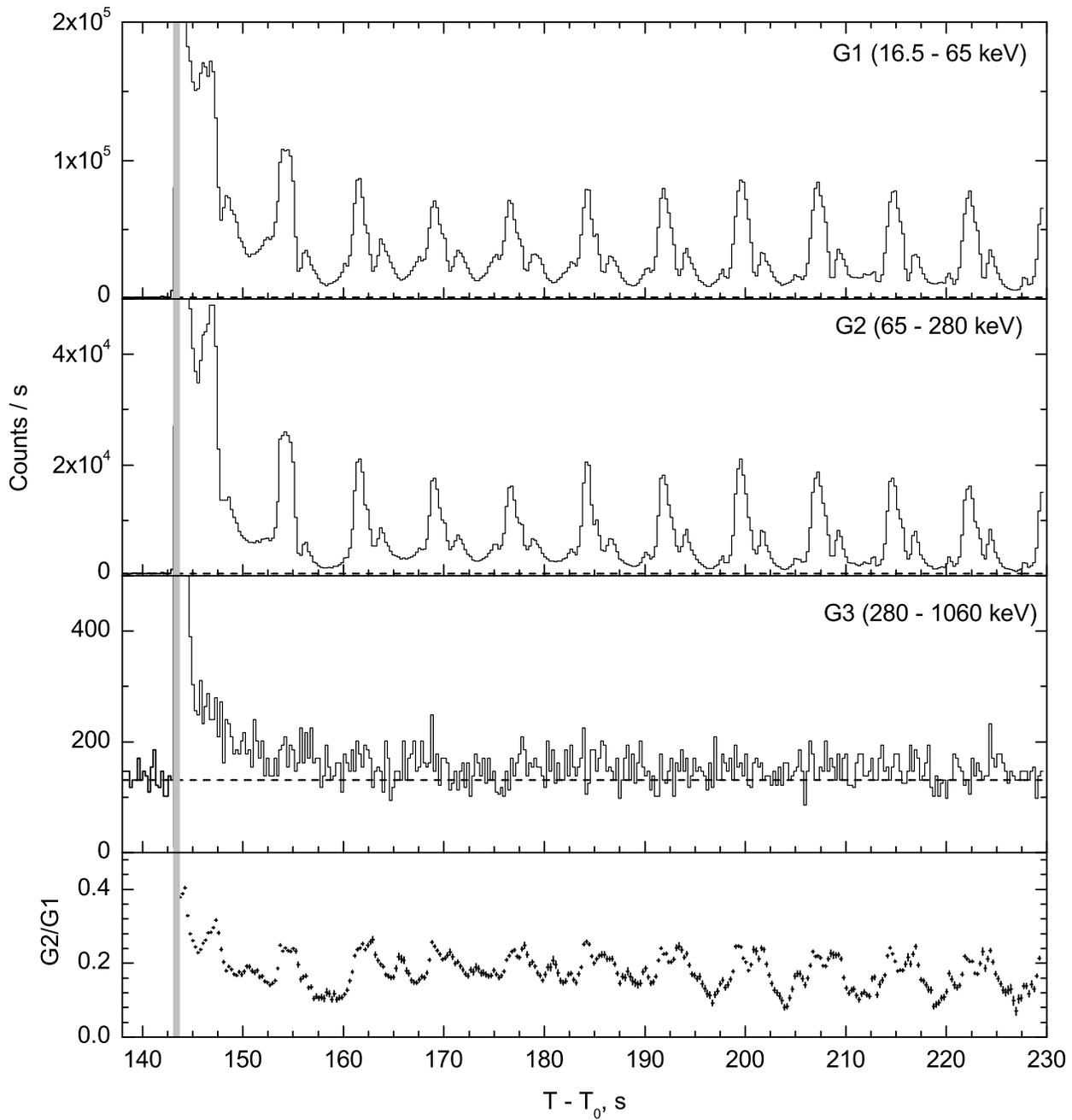}
\caption{Time history of the giant flare on December 27, 2004,
recorded by the Konus-Wind detector, T$_0$=21:27:59.447 UT. The
trigger part of the time history in three windows: G1(16.5--65~keV),
G2(65--280~keV), and G3(280--1060~keV), and the hardness ratio
G2/G1. The vertical gray line marks the time interval when the
detector was in a state of complete saturation. \label{tKW_giant}}
\end{figure}
\clearpage
\begin{figure}
\centering
\includegraphics[width=\textwidth]{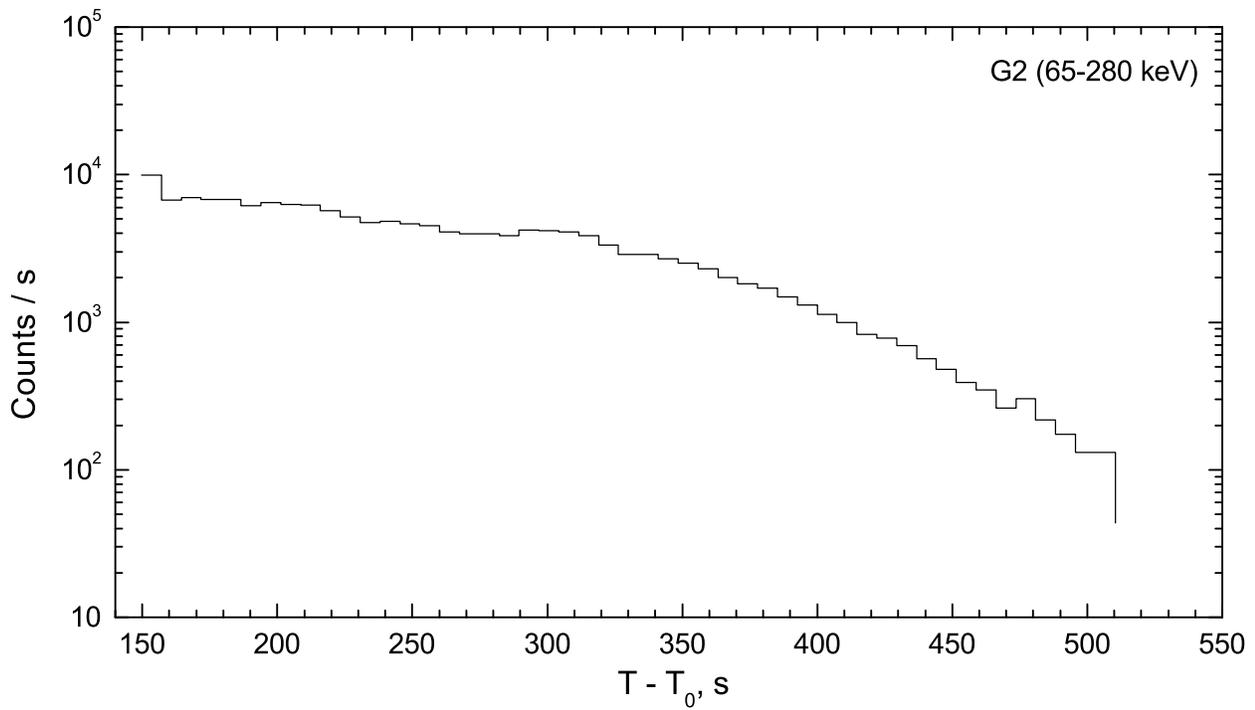}
\caption{Time history of the tail of the giant flare on December 27,
2004, in the G2 window with a resolution of 7.36~s. Konus-Wind data,
T$_0$=21:27:59.447 UT. The tail rapidly decays after T-T$_0$=500~s.
\label{tKW_giant_G2_averaged}}
\end{figure}
\clearpage
\begin{figure}
\centering
\includegraphics[width=\textwidth]{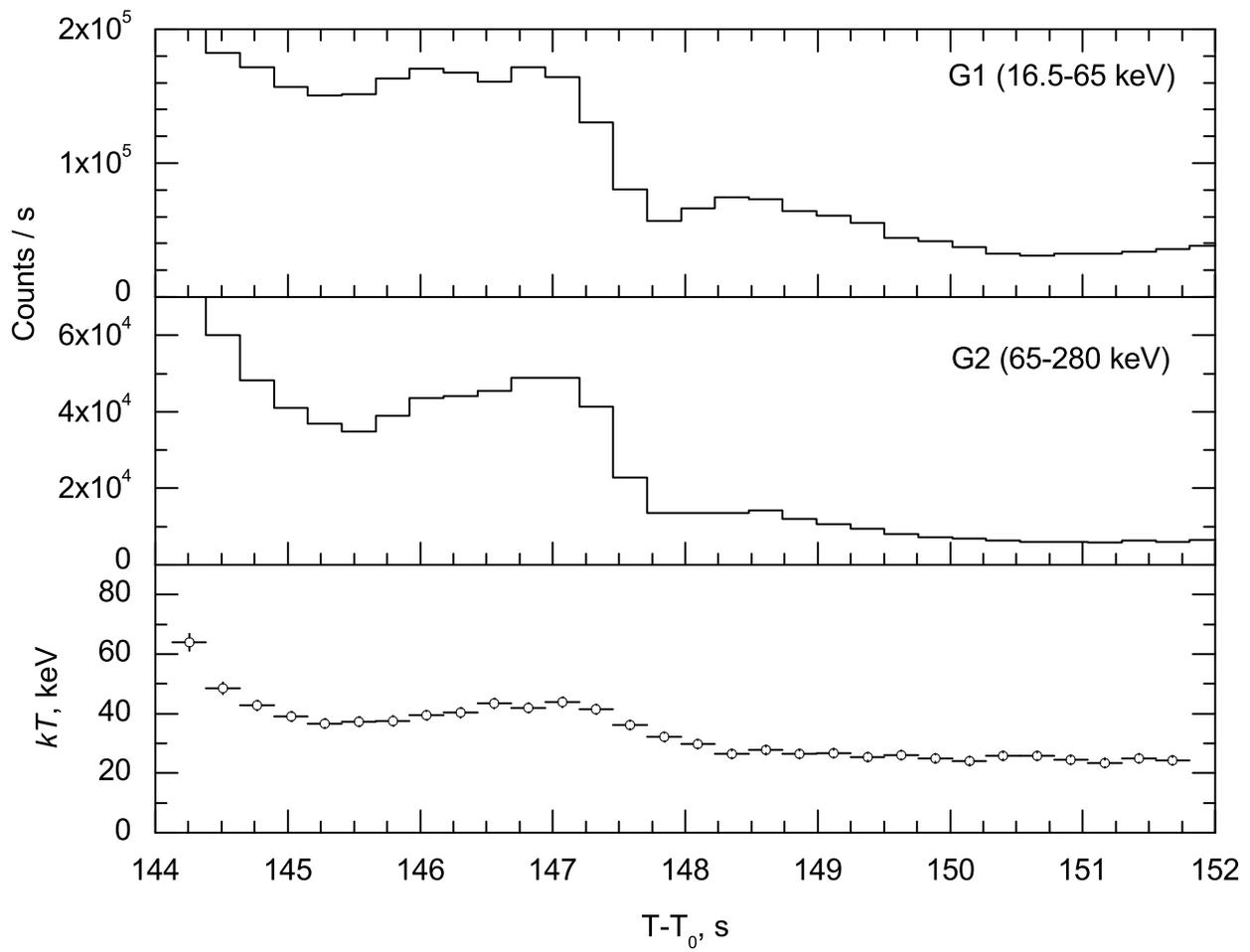}
\caption{Initial part of the tail of the giant flare on December 27,
2004. \label{tKW_giant_tail_start}}
\end{figure}
\clearpage
\begin{figure}
\centering
\includegraphics{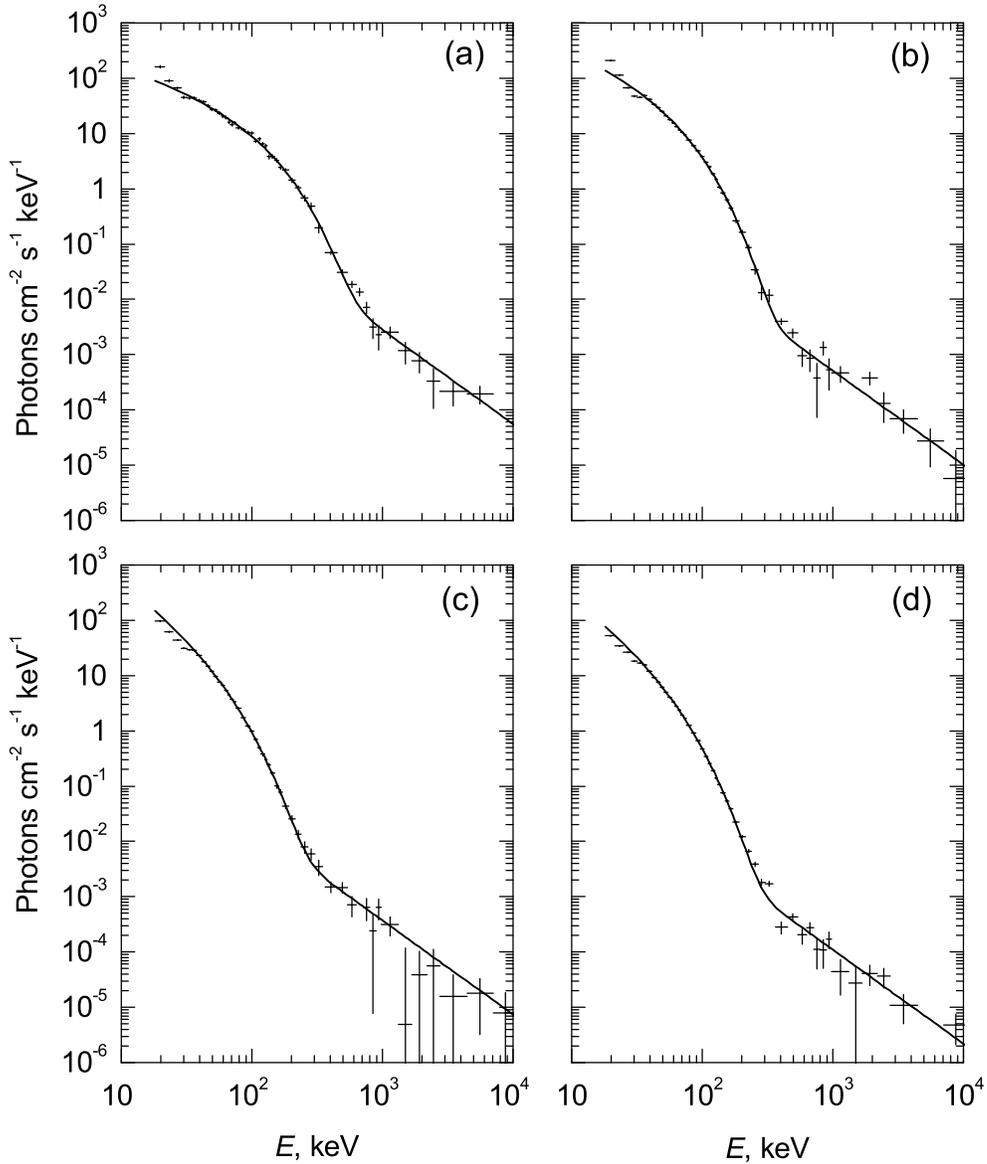}
\caption{Spectra of the pulsating tail measured by the Konus-Wind
detector. All spectra consist of two components: a low-energy
component similar to the spectra of recurrent bursts with $kT \simeq
30$~keV and a hard power-law component with an index $\gamma \simeq
-1.7$. For each spectrum, this two-component model is indicated by
the solid line. The spectra were measured in the following
intervals: (a) 0.872--1.384~s, (b) 1.384--4.456~s, (c)
4.456--8.808~s, and (d) 8.808--74.344~s (the time is measured from
the onset of the giant flare, T-T$_0$=142.98~s).
\label{spKW_giant_tail}}
\end{figure}
\clearpage
\begin{figure}
\centering
\includegraphics{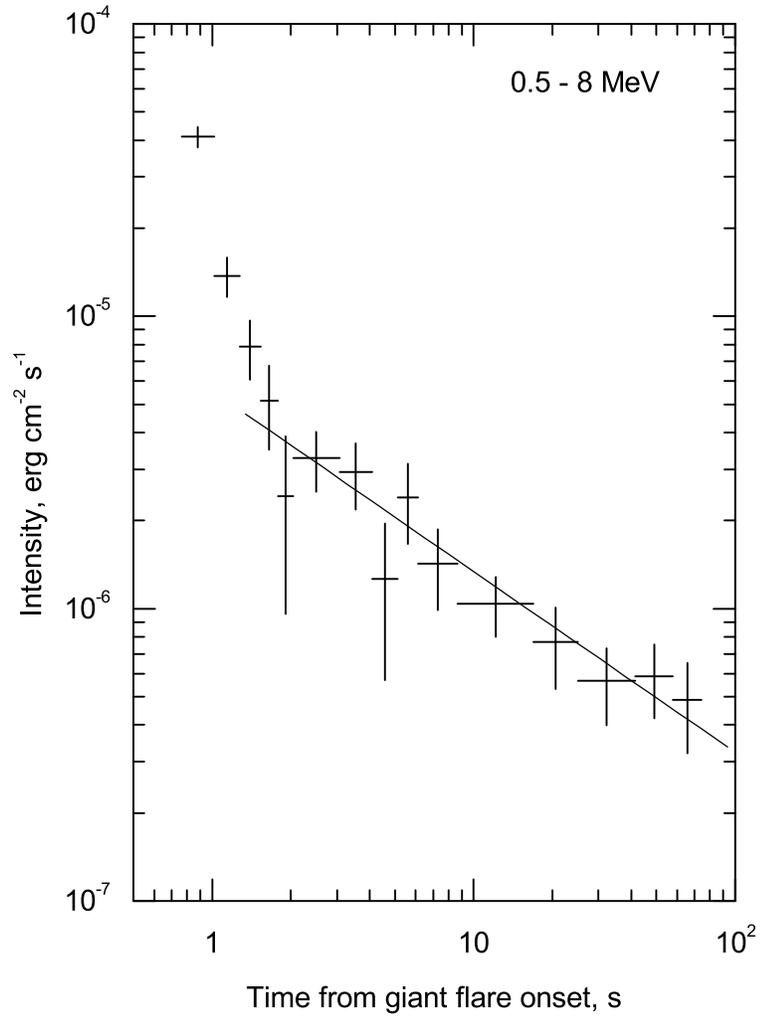}
\caption{Intensity of the hard (0.5--8~MeV) component vs. time (the
time is measured from the onset of the giant flare,
T-T$_0$=142.98~s). The solid line indicates the dependence $\propto
t^{-0.6}$. \label{tKW_giant_hard_tail}}
\end{figure}
\clearpage
\begin{figure}
\centering
\includegraphics[width=\textwidth]{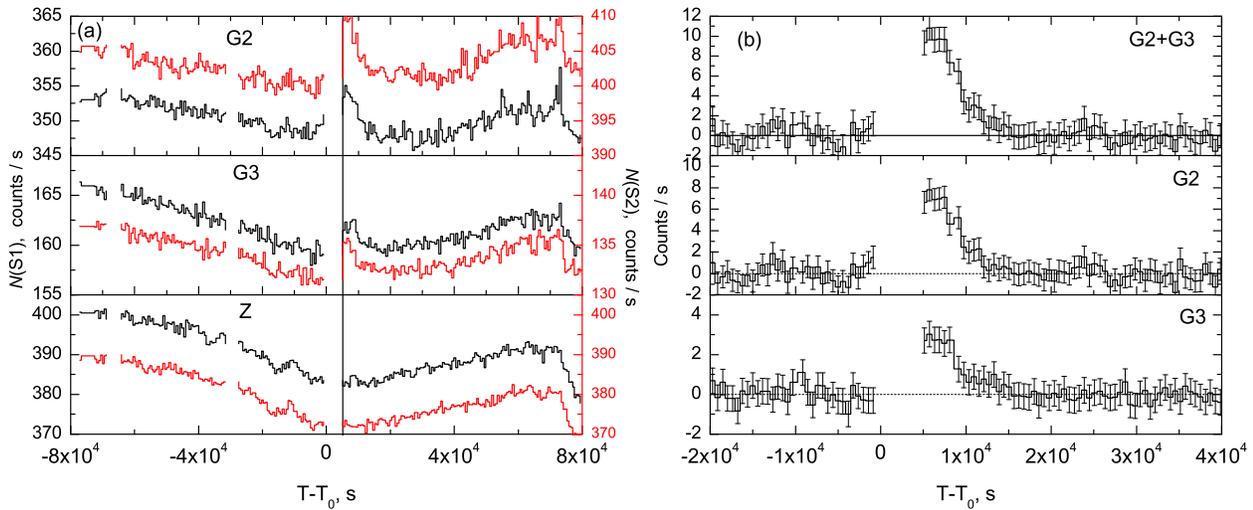}
\caption{Record of the background summed over 800~s in the G2, G3,
and Z windows. T$_0$ corresponds to the Konus-Wind trigger time. (a)
The data recorded on December 27--28, 2004. The count rates recorded
by the S1 (dark curve) and S2 (light curve) detectors are given for
each window. The gaps in the record correspond to the data output
intervals after the completion of the trigger record. The vertical
bar marks the record resumption time after the output of trigger
data on the giant flare. (b) A segment of the record near the giant
flare -- the record in the G2 and G3 windows, the count rates were
averaged over the two detectors (the background was subtracted, 1$
\sigma$ errors of the measured count rates are given).
\label{tKW_afterglow}}
\end{figure}
\clearpage
\begin{figure}
\centering
\includegraphics{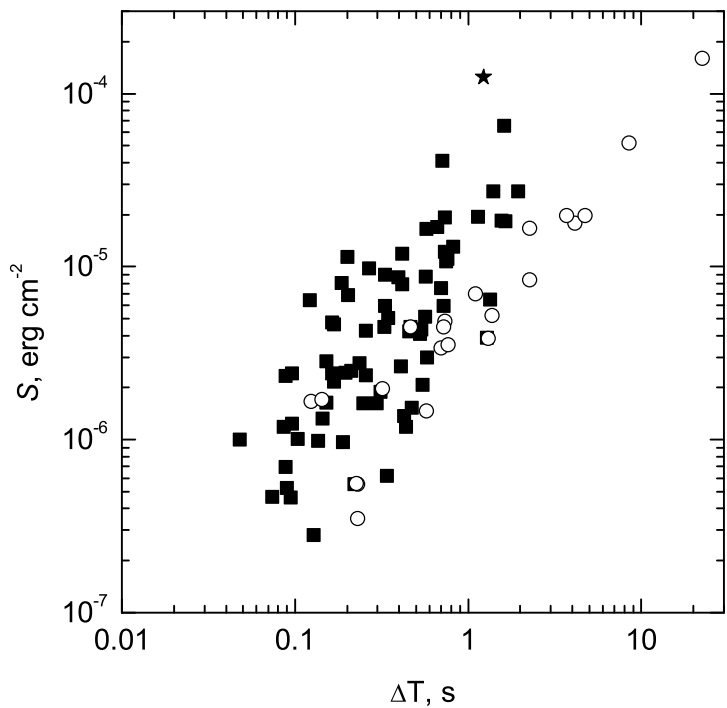}
\caption{$S$--$\Delta$T diagram for the trigger recurrent bursts
detected by Konus-Wind and Helicon-Coronas-F before (filled squares;
72 bursts) and after (open circles; 22 bursts) the giant flare on
December 27, 2004 (the series of bursts are not included). The
asterisk marks the precursor of the giant flare.
\label{Fluence_Duration}}
\end{figure}
\clearpage
\begin{figure}
\centering
\includegraphics[height=0.9\textheight]{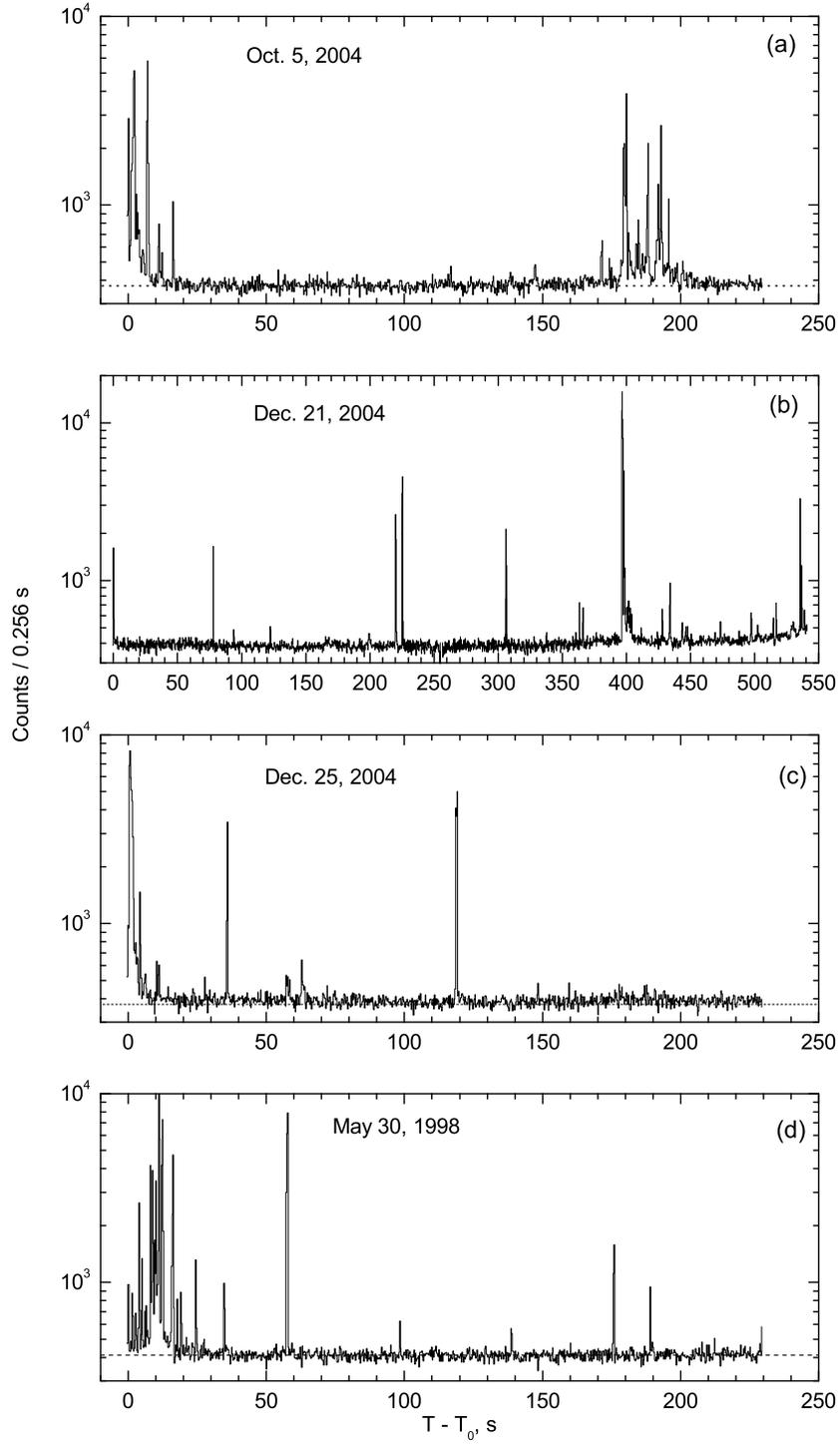}
\caption{Time histories of the series of bursts in SGR~1806-20 (a,
b, c) and SGR~1900+14 (d). The fluence in the range 20--200 keV is
$7.6 \times 10^{-5}$ (a), $3.3 \times 10^{-5}$ (b), $7.6 \times
10^{-5}$ (c), and $5.9 \times 10^{-5}$~erg~cm$^{-2}$ (d).
\label{tBurstSeries}}
\end{figure}
\clearpage
\begin{figure}
\centering
\includegraphics[]{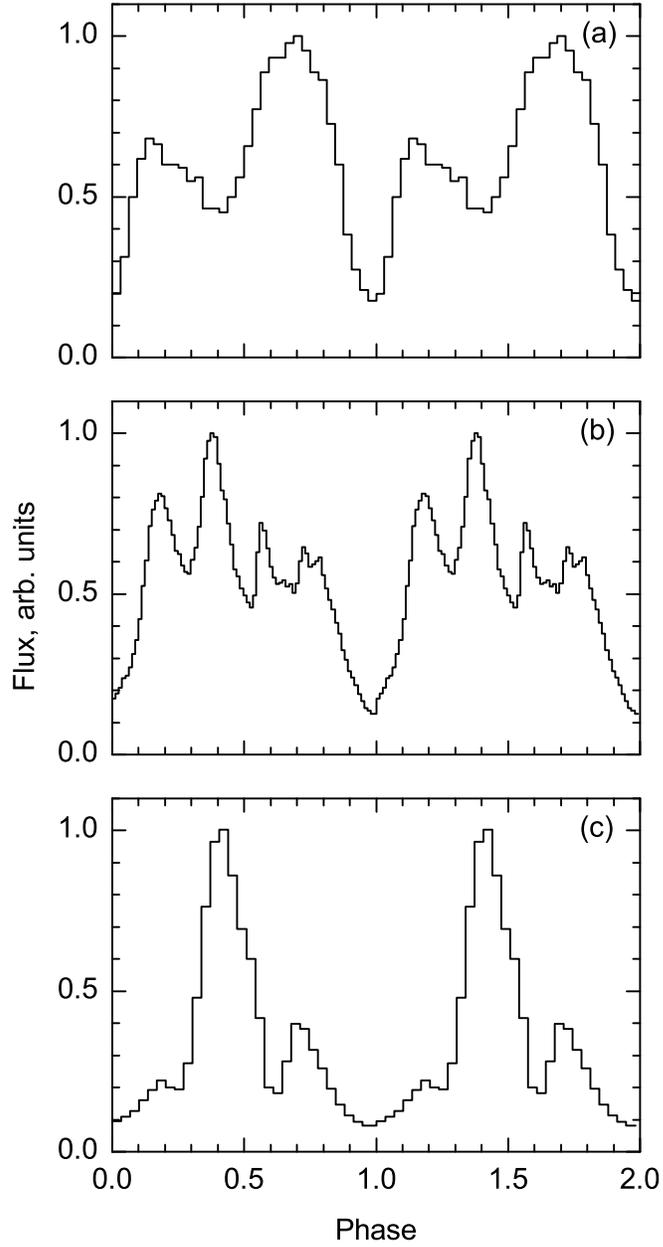}
\caption{Phase-averaged profiles of the pulsating tails of the three
giant flares from SGRs: (a) on March 5, 1979, from SGR~0526-66
(Venera-11 and Venera-12 data), (b) on August 27, 1998, from
SGR~1900+14 (Konus-Wind data), and (c) on December 27, 2004, from
SGR~1806-20 (Konus-Wind data). \label{tGiantTailsFolded}}
\end{figure}
\end{document}